\documentclass{iaus}
\usepackage{graphicx}

\title[Numerical simulations of galaxy evolution] 
{Numerical simulations of galaxy evolution in cosmological context}

\author[M. Martig, F. Bournaud \& R. Teyssier]   
{Marie Martig$^1$$^,$$^2$, Fr{\'e}d{\'e}ric Bournaud$^1$$^,$$^2$
 \and Romain Teyssier$^1$$^,$$^2$}

\affiliation{$^1$CEA, IRFU, SAp. F-91191 Gif-sur-Yvette,France.\\[\affilskip]$^2$Laboratoire AIM, CNRS, CEA/DSM, Universit{\'e} Paris Diderot. F-91191 Gif-sur-Yvette, France.}
\pubyear{2008}
\volume{254}  
\pagerange{--}
\jname{The Galaxy Disk in Cosmological Context}
\editors{J. Anderson, J. Bland-Hawthorn \& B. Nordstrom, eds.}
\begin{document}

\maketitle

\begin{abstract}
Large volume cosmological simulations succeed in reproducing the large-scale structure of the Universe. However, they lack resolution and may not take into account all relevant physical processes to test if the detail properties of galaxies can be explained by the CDM paradigm. On the other hand, galaxy-scale simulations could resolve this in a robust way but do not usually include a realistic cosmological context.

To study galaxy evolution in cosmological context, we use a new  method that consists in coupling cosmological simulations and galactic scale simulations. For this, we record merger and gas accretion histories from cosmological simulations and re-simulate at very high resolution the evolution of baryons and dark matter within the virial radius of a target galaxy. This allows us for example to better take into account gas evolution and associated star formation, to finely study the internal evolution of galaxies and their disks in a realistic cosmological context.

We aim at obtaining a statistical view on galaxy evolution from z $\simeq$ 2 to 0, and we present here the first results of the study: we mainly stress the importance of taking into account gas accretion along filaments to understand galaxy evolution.
\end{abstract}

\firstsection 
\section{Introduction}

The morphology of galaxies in the Local Universe is well constrained by observations, but is still largely unexplained. Indeed, large volume cosmological simulations fail to reproduce realistic galaxies. For instance, the disks formed are often too concentrated : it is the ``angular momentum problem'', well known since the early work of \cite{Navarro1991}.  It is still unclear whether this is an intrinsic problem of the $\Lambda$CDM paradigm or if something (i.e. resolution, physical processes...) is missing in these simulations. 

Another puzzle is the question of disk survival till z=0 (\cite{Koda2007}). For instance, \cite{Kautsch2006} study a large sample of edge-on spiral galaxies in the SDSS and find that a significant fraction of them (i.e. roughly one third) are bulgeless or ``superthin''. This is still unexplained by cosmological models. Indeed, $\Lambda$CDM predicts that galaxy interactions are frequent (see e.g. the recent work by \cite{Stewart2007}). More exactly, major mergers, that are well known to destroy disks to form ellipticals (\cite{Barnes1991}) are rather rare, but minor mergers are much more common. These minor mergers can thicken disks, and if frequent enough could even form elliptical galaxies (\cite{Bournaud2007}). The problem is then to find whether $\Lambda$CDM predicts too many mergers, or if the satellites have properties and orbital parameters such that they have little influence on the galactic disks. Also, gas accretion along filaments could fuel a thin disk and counteract the effect of mergers (\cite{Dekel2005}, \cite{Keres2005}, \cite{Ocvirk2008}).

To study the properties of galaxies at low and high redshift, it thus seems necessary to take the full cosmological context into account. Large scale cosmological simulations could of course achieve this goal and give a statistical view on galaxies at each redshift, but for now they mainly lack resolution at the galactic scale. On the contrary, small volume cosmological simulations like the one performed by \cite{Naab2007} can resolve galactic scales in detail but are so time-consuming that obtaining a statistical sample is for now a challenge.

A first method to solve these problem is to use semi-analytical models, i.e. extracting merger trees from cosmological simulations and using different recipes to infer physical properties of galaxies (\cite{Somerville2001}, \cite{Hatton2003}, \cite{Khochfar2005}). The drawback is that approximations are necessary.

Another possibility has been explored by \cite{Kazantzidis2007}, \cite{Read2007} and \cite{Villalobos2008} : they extract merger histories from cosmological simulations and  re-simulate these histories at higher resolution. Nevertheless, they perform collisionless simulations with no gas component, neither in the main galaxy, nor in satellites, nor in filaments. 

We here present a new approach where we re-simulate at high resolution a history given by a cosmological simulation, using self consistent realistic galaxies (the main galaxy and the satellites have a gas disk, a stellar disk and a dark matter halo), and we also take into account gas accretion from cosmic filaments. Our goal is to obtain a statistical sample of merger and accretion histories in a $\Lambda$CDM context to simulate the resulting galaxies and to compare our results to observations at various redshifts. 

After a description of the technique used, we will present our first results and emphasize the importance of gas accretion along filaments to understand galaxy evolution.
\section{Method}

\subsection{Analysis of the cosmological simulation}
Merger histories and accretion data are extracted from a dark matter only cosmological simulation performed with the AMR code RAMSES (\cite{Teyssier2002}). This simulation has an effective resolution of 512$^3$ and a comoving box length of 20 h$^{-1}$ Mpc. The mass resolution is 6.9$\times$10$^6$ M$_{\odot}$, so that a Milky Way type halo is made of a few 10$^{5}$ particles. The cosmology is set to $\Lambda$CDM with $\Omega_m$=0.3, $\Omega_{\Lambda}$=0.7, H$_0$=70 km.s$^{-1}$.Mpc$^{-1}$ and $\sigma_8$=0.9.

In this simulation, halos are detected with the HOP algorithm (\cite{Eisenstein1998}), with $\delta_{\rm peak}$=240, $\delta_{\rm saddle}$=200 and $\delta_{\rm outer}$=80 (the minimal number of particles per halo is fixed to 10). In the following, we also take into account particles that do not belong to a halo, and we consider them as diffuse accretion.

The halo of which we want to build the merger and accretion history is then chosen in the final snapshot of the simulation (at z = 0) and is traced back to higher redshift (typically z $\simeq$ 2) : we will call it the main halo. From z $\simeq$  2 to z = 0, each halo or particle (in the case of diffuse accretion) entering a sphere around the main halo (the radius of this sphere is the virial radius of the main halo at z=0) is recorded, with its mass, position, velocity and spin (spin is of course omitted for diffuse accretion). 

\subsection{High resolution re-simulation}
\subsubsection{The PM code}
The history that has been extracted from the cosmological simulation is re-simulated with a particle-mesh code (\cite{BC02}).
Gas dynamics is modeled with a sticky-particle scheme with $\beta_r$=0.8 and $\beta_t$=0.7, and star formation is computed according to a Kennicutt law with an exponent 1.5.

The maximum spatial resolution is 130 pc. For the two simulations shown hereafter, the mass resolution varies from 1.2$\times$10$^4$ M$_{\odot}$ to 2.1$\times$10$^4$ M$_{\odot}$ for gas particles, from 6$\times$10$^4$ M$_{\odot}$ to 1.4$\times$10$^5$ M$_{\odot}$ for stellar particles and from 1.2$\times$10$^5$ M$_{\odot}$ to 4.4$\times$10$^5$ M$_{\odot}$ for dark matter particles. This allows to have a total number of particles of the order of 15$\times$10$^6$ at the end of both simulations.

\subsubsection{Model galaxies}

Each halo of the cosmological simulation (i.e. the main halo as well as all the interacting satellites) is replaced with a realistic galaxy, having a disk, a bulge and of course a dark matter halo. The total mass of the galaxy is divided in 20\% of baryons and 80\% of dark matter (the mass of dark matter being given by the cosmological simulation). The dark matter halo follows a Burkert profile extended to its virial radius, with a core radius chosen to follow the scaling relations given in \cite{Salucci2000}. The disk radius of each galaxy is proportional to the square root of its mass so that the surface density is constant from one galaxy to another.

The gas fraction in the disk is 30\% for galaxies that have a halo mass lower than 10$^{11}$~M$_{\odot}$. For galaxies that have a greater halo mass, the gas fraction is set to 30\% at high redshift (z$>$0.8) and 15\% at low redshift.

\begin{figure}
\begin{center}
\includegraphics[width=6.5cm]{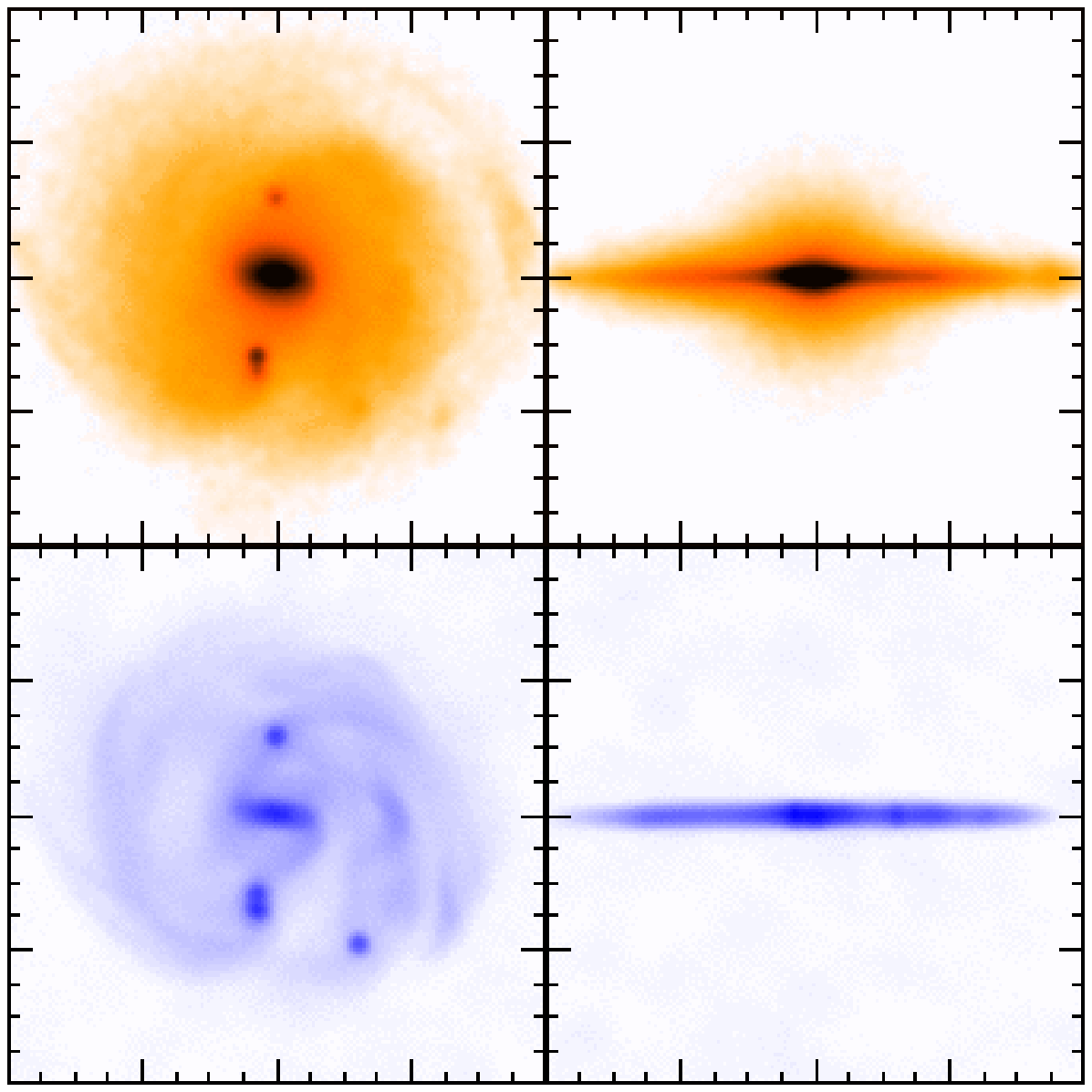}
\includegraphics[width=6.5cm]{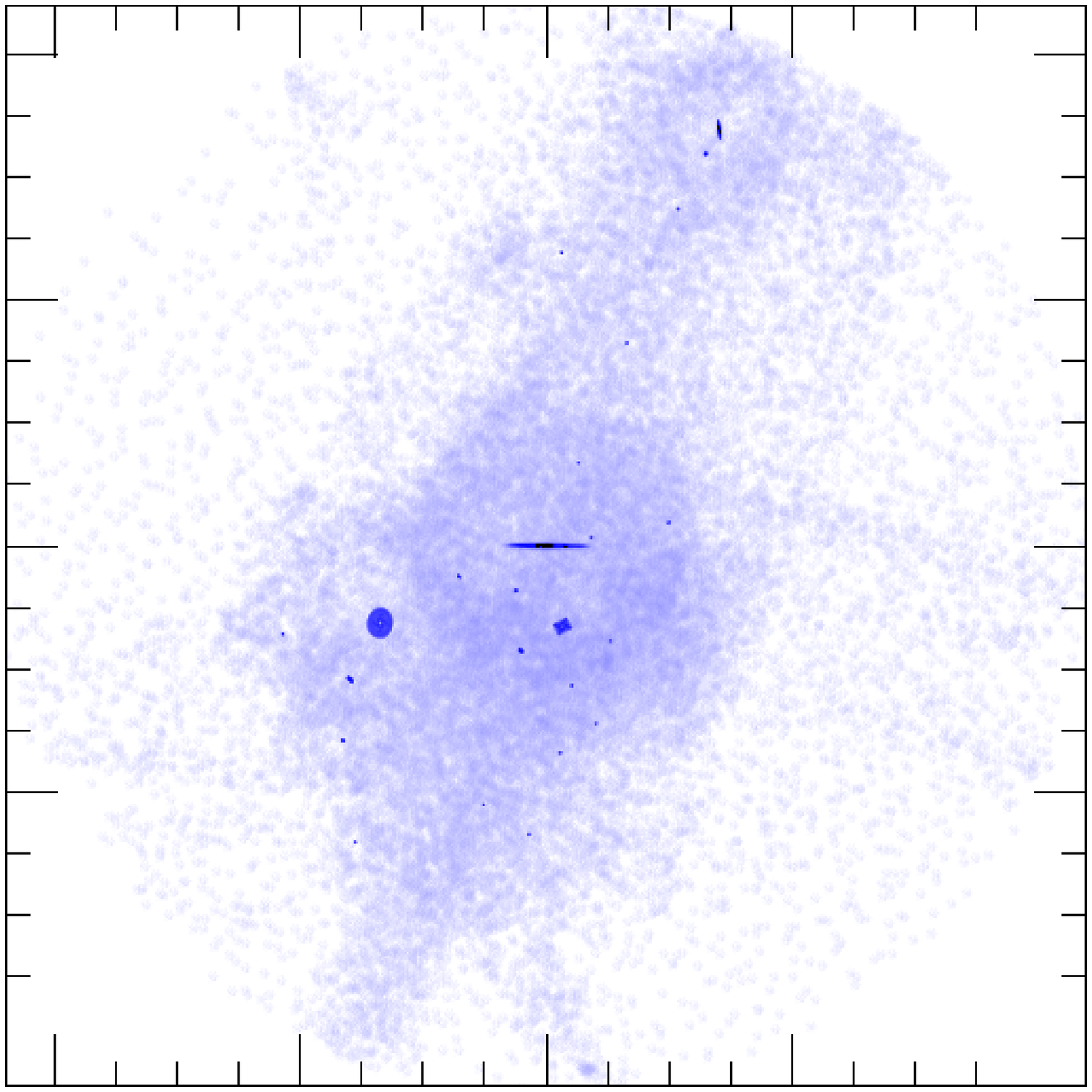}
\end{center}
\caption{Left : Initial distribution of stars (top panel) and gas (bottom panel) for the main galaxy, seen face-on and edge-on (each panel is 40 kpc x 40 kpc in size). Right : large scale view of the gas distribution in a simulation box (the panel is 440 kpc x 440 kpc in size).}\label{init}
\end{figure}

Figure \ref{init} (left side) shows for example the initial distribution of gas and stars in the main galaxy.

\subsubsection{Diffuse accretion}
Each dark matter particle that is considered as diffuse accretion in the cosmological simulation is replaced with a small blob of particles, containing in mass 20\% of gas and 80\% of dark matter. 

The right side of figure \ref{init} shows an example of simulation where the main galaxy (edge-on) is surrounded by accreted gas (clearly in a filament) and a few satellite galaxies.
\subsection{Two examples}
We present here the first results concerning two simulations, that have been chosen to have a mass at z=0 of the order of magnitude of the mass of the Milky Way. They have very different histories. 

\begin{figure}
\begin{center}
\includegraphics[width=6.2cm]{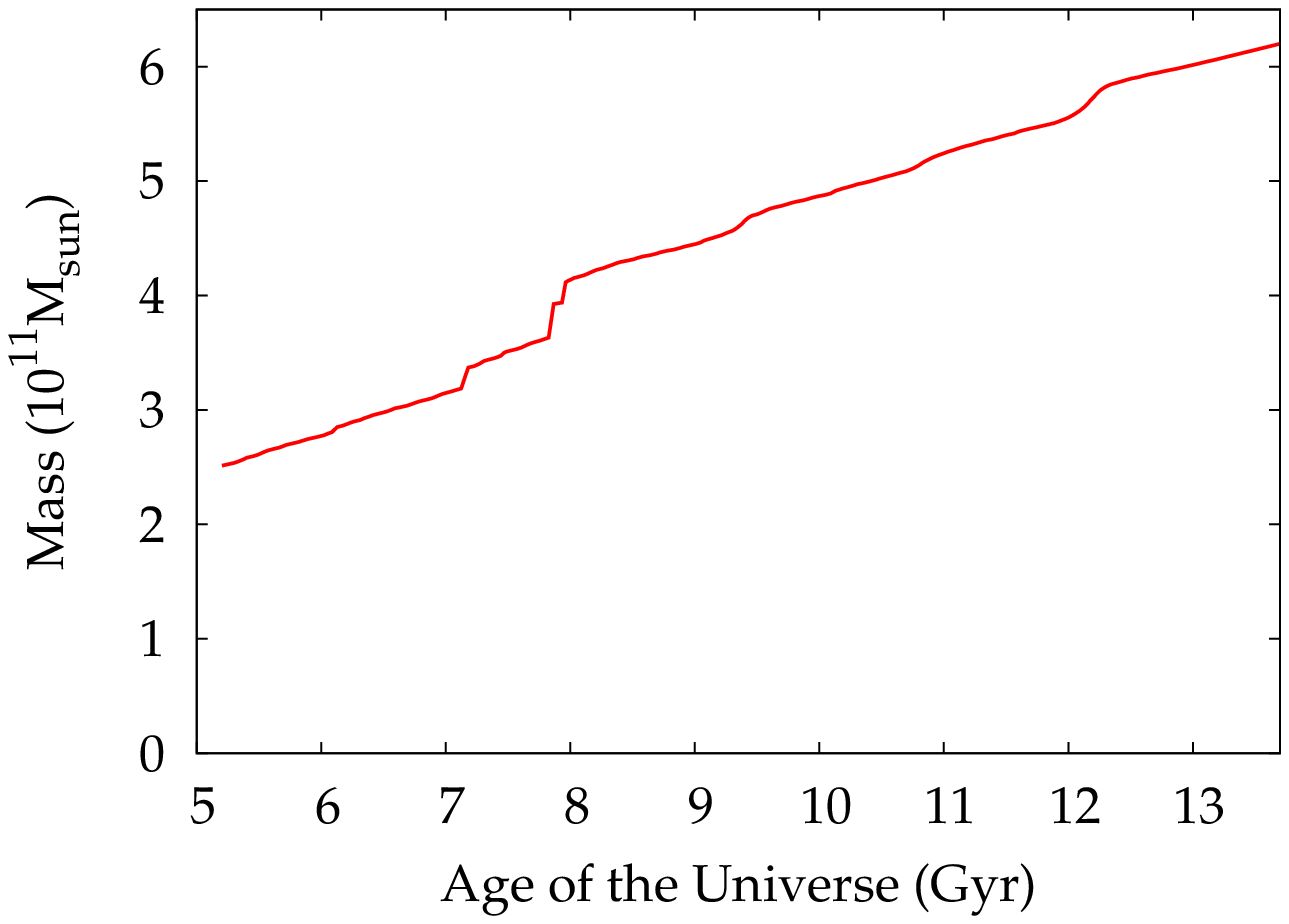}
\includegraphics[width=6.2cm]{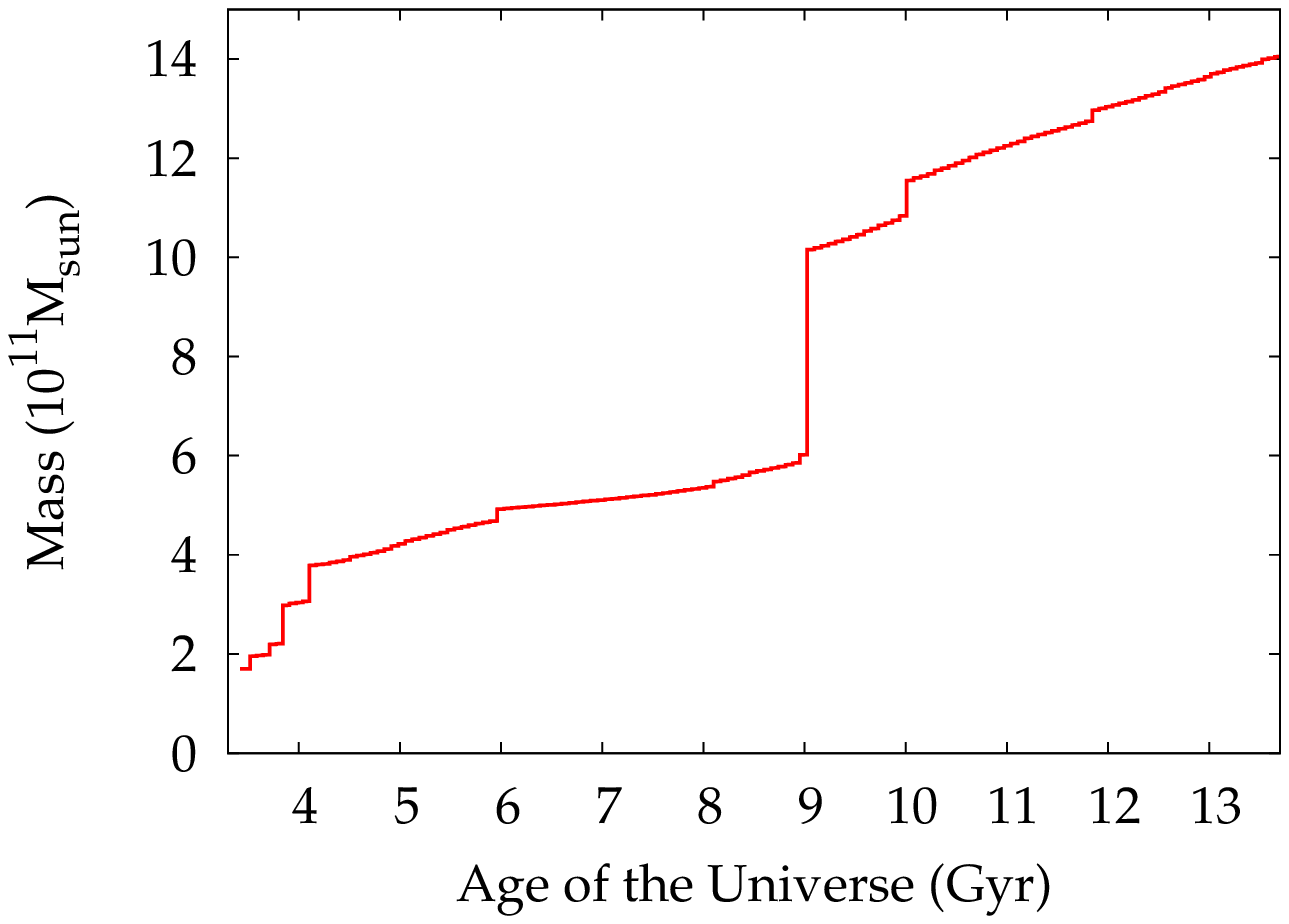}
\end{center}
\caption{Evolution of the total mass of dark matter in the simulation box as a function of time for the two simulations studied here : in the left case, the mass growth is dominated by accretion, and in the right one by mergers}\label{history}
\end{figure}

In the the first one, the mass growth of the galaxy is dominated by diffuse accretion (at a mean rate of $\sim$ 5 M$_{\odot}	$/yr). Only some very minor mergers take place, the most important of these mergers having a mass ratio of 12:1 (see on the left panel of figure \ref{history} the mass evolution as a function of time). We will call this simulation \textit{``the calm case''}.

The other simulation also contains diffuse accretion, but is mainly dominated by mergers. There is a first period of repeated minor and major mergers (mass ratios 8:1, 10:1, 3:1 and 4:1) at the very beginning of the simulation, then a calm phase and finally a major merger (mass ratio 1.5:1) at low redshift (see right panel of figure \ref{history}). We will call it \textit{``the violent case''}.

\section{Results}
\subsection{The calm case}

\begin{figure}
\begin{center}
\includegraphics[width=12cm]{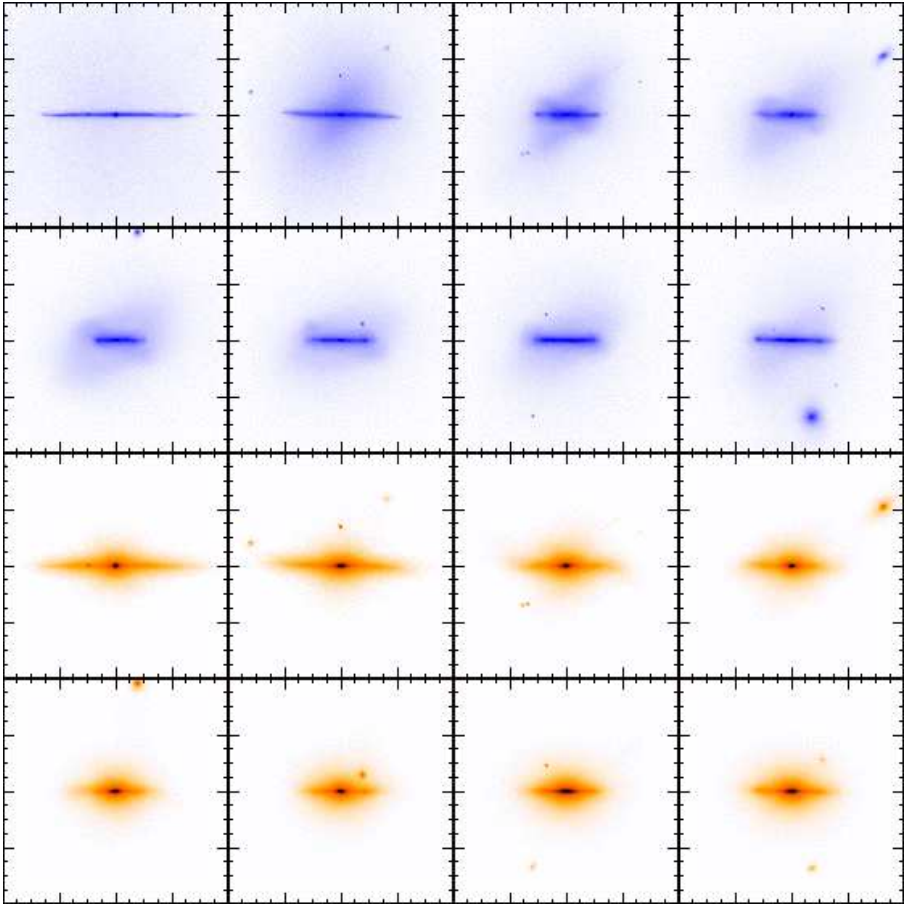}
\end{center}
\caption{Evolution of the distribution of gas (top panels) and stars (bottom panels) for the calm case. Snapshots are taken every Gyr and each panel is 40 kpc x 40 kpc in size.}\label{evol_70}
\end{figure}

The evolution of the distribution of gas and stars is shown in figure \ref{evol_70}. Gas is smoothly accreted around the galaxy and falls onto the disk. Minor mergers are not strong and frequent enough to destroy the stellar disk. They only slightly heat it, and a thin stellar disk is rebuilt thanks to gas from diffuse accretion along the filaments.
The thin disk is mainly formed from stars younger than 4 Gyr, and has a well-defined structure with two spiral arms.

\subsection{The violent case}

\begin{figure}
\begin{center}
\includegraphics[width=12.8cm]{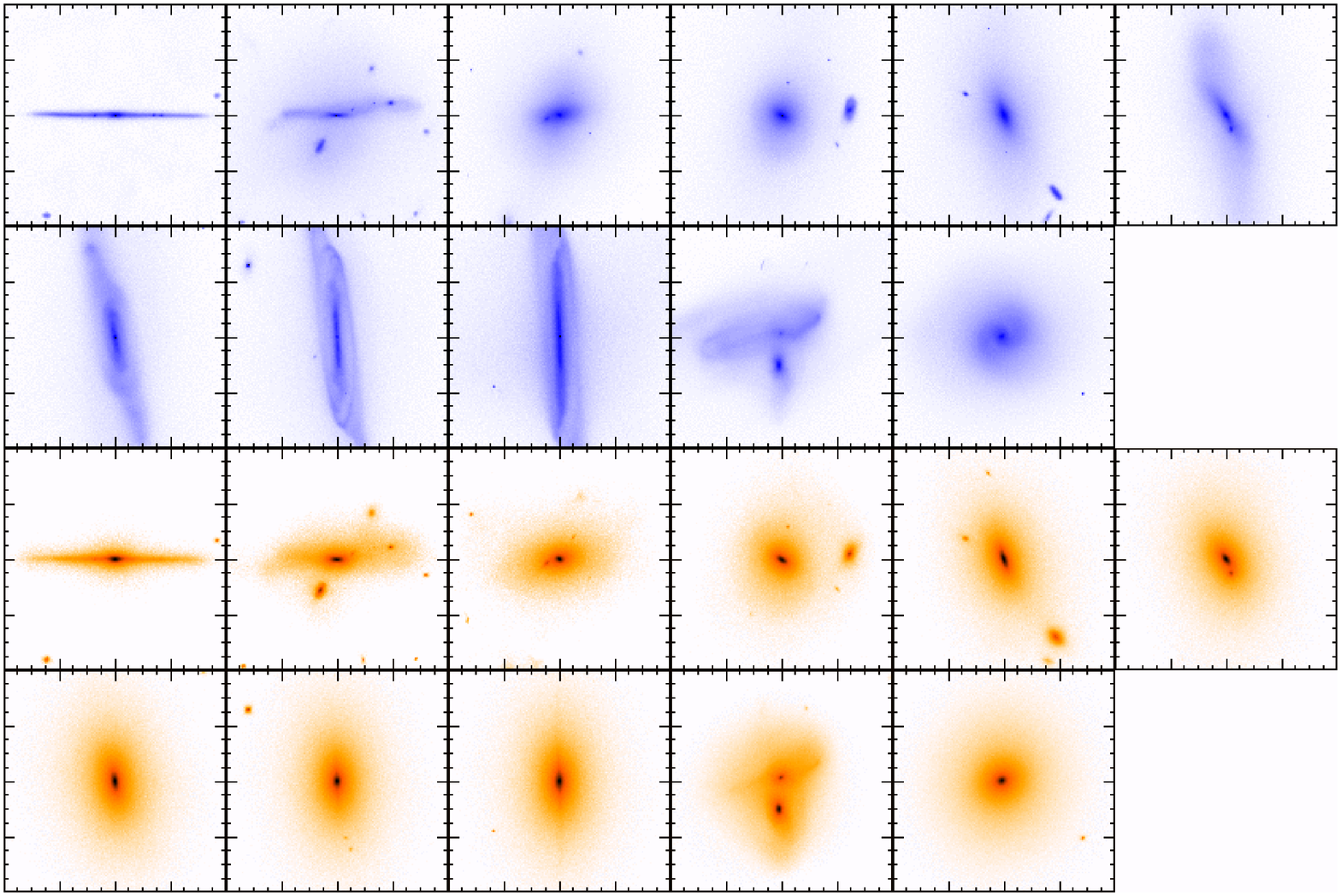}
\end{center}
\caption{Evolution of the distribution of gas (top panels) and stars (bottom panels) for the violent case. Snapshots are taken every Gyr and each panel is 40 kpc x 40 kpc in size.}\label{evol_35}
\end{figure}

In this case, the evolution of the morphology of the galaxy is totally different (see figure \ref{evol_35}). The disk is destroyed early by the first series of mergers. In fact, after the first of these mergers (which has a mass ratio of 8:1) the disk is already very perturbed, and the following mergers contribute to the transformation of the galaxy into an elliptical.

\begin{figure}
\begin{center}
\includegraphics[width=6cm]{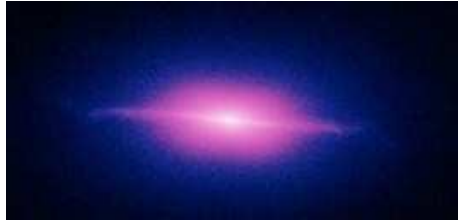}
\end{center}
\caption{Projected stellar mass density at z = 0.2 for the violent case.}\label{disk}
\end{figure}

Nevertheless, thanks to gas accretion that takes place along a filament, a gas disk is gradually re-built into the elliptical galaxy (this would not happen if only mergers were taken into account in the simulation). New stars form in this disk, forming a young stellar disk inside the old spheroid (see figure \ref{disk}), this disk being in a perpendicular plane with respect to the initial disk. Finally, the last major merger (with a mass ratio of 1.5:1) destroys this disk and the galaxy becomes elliptical again.

\section{Conclusion}
In order to study galaxy evolution in cosmological context, we have successfully developed a technique that allows us to perform high resolution simulations taking into account realistic merger and gas accretion histories. 

The first two simulations shown here do not allow us to draw any general conclusion on galaxy evolution in a $\Lambda$CDM context. Nevertheless, we can already confirm that even low mass satellites can thicken disks and that ellipticals from both through repeated minor mergers and major mergers. We also emphasize that gas accretion from filaments can allow to rebuild a thin disk in a galaxy, which proves the absolute necessity to take this accretion into account to understand galaxy evolution.

\end{document}